# Scattering of a plane wave by an inhomogeneous 1D dielectric layer with gradient refractive index

N.A. Vanyushkin, A.H. Gevorgyan, S.S. Golik

Far Eastern Federal University, 10 Ajax Bay, Russky Island, Vladivostok, 690922, Russia,

**Abstract**

We propose a new method for calculating reflection and transmission coefficients for an arbitrarily polarized electromagnetic plane wave incident on a one-dimensional dielectric medium of finite thickness and with dielectric permittivity being an arbitrary continuous function of the coordinate. We have shown that the problem of plane wave scattering by an inhomogeneous layer is reduced to a system of first order differential equations that contain the derivative of the refractive index or dielectric permittivity of the layer, which can be used, for example, when searching for an analytical solution. This method also makes it easy to obtain the distribution of the field strength within the layer. The reflection spectra and field distribution obtained using this method were compared with the analytical solution based on Mathieu functions.

1. **Introduction**

The problem of electromagnetic wave propagation in a one-dimensional isotropic medium with an arbitrary or random distribution of the refractive index has practical applications in many fields. It can be solved using many methods, such as classical methods for solving ordinary differential equations, integral equations method, transfer matrix and scattering matrix methods, Green's function method, invariant immersion method, phase function method, semiclassical method, etc. [1–12]. Nevertheless, various new methods of solving this problem continue to be developed in recent years [13-17]. Each of these methods has its advantages and disadvantages. The choice of one or another method is usually determined by the problem formulation, namely, which aspect of the problem is of the greatest interest, the computational complexity, and the possibility of obtaining analytical expressions. Usually, the problem of electromagnetic wave propagation through a layered medium is reduced to a linear boundary problem for the field amplitude inside the medium. However, if only the amplitudes of the transmitted and reflected waves are to be found, then the problem can be reduced to the Cauchy problem for differential equations for transmission $T$ and reflection $R$ coefficients.

In [18,19] a method for solving this system of equations by a suitable choice of two functions which are combinations of coefficients $T$ and $R$ was proposed and generalized. The main advantage of this method as compared to standard methods from a computational point of view is that the solution of the system of two linear first order differential equations is easier than the wave equation. Note that there is a similar method proposed by Bovard [20], which is also based on solving a system of two coupled linear differential equations. The purpose of this paper is to continue the development of the method developed by [18], namely, the problem of scattering of an electromagnetic wave by an inhomogeneous layer taking into account the spatial gradient of the refractive index will be considered.

2. **Problem statement and solution**

Let us consider the reflection and transmission of a plane electromagnetic wave through an inhomogeneous layer (Fig. 1), which is located between the planes $z=0$ and $z=L$, and its permittivity $\varepsilon$ is an arbitrary continuous function of $z$ only. We assume that the medium is isotropic, nonmagnetic ($\mu = 1$), and does not absorb. Let us also assume that the incidence plane coincides with the plane $(x, z)$, and the wave is incident at an angle $\alpha$ to the normal of the layer boundary, which coincides with the plane $(x, y)$. The regions $z < 0$ and $z > L$ are filled with homogeneous non-absorbing dielectrics with $\varepsilon_0 = \varepsilon(0)$ and $\varepsilon_L = \varepsilon(L)$, respectively. Thus, the dielectric permittivity in our problem is continuous at the layer boundaries.

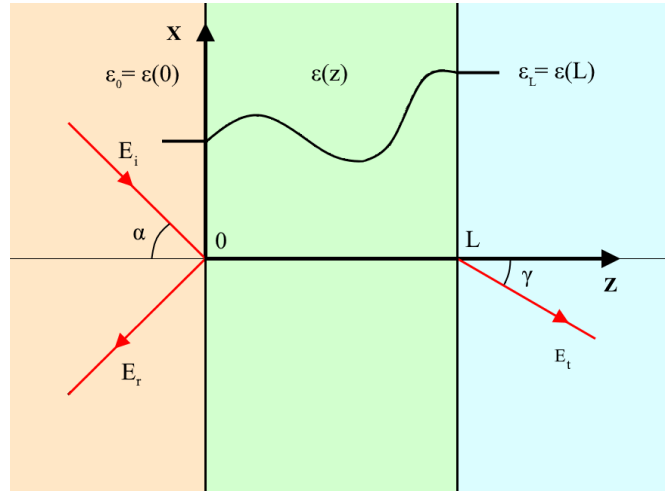

Figure 1. Transmission of a plane wave incident on a one-dimensional isotropic dielectric medium with an arbitrary dielectric permittivity

The amplitudes for the electric field of the incident, reflected and transmitted plane wave will be denoted by $\mathbf{E}_i$, $\mathbf{E}_r$ and $\mathbf{E}_t$, respectively. These fields expand as:

$$\mathbf{E}_{i,r,t} = E^p_{i,r,t}\mathbf{n}_p + E^s_{i,r,t}\mathbf{n}_s = \begin{pmatrix} E^p_{i,r,t} \\ E^s_{i,r,t} \end{pmatrix} \tag{1}$$

where $\mathbf{n}_p$ and $\mathbf{n}_s$ are the unit vectors of p- and s- polarization, $E^p_{i,r,t}$ and $E^s_{i,r,t}$ are the corresponding amplitudes of the incident, reflected, and transmitted waves. The complex amplitudes of transmission and reflection for s- and p-waves can be written in the form:

$$R^{s,p} = \frac{E^{s,p}_r}{E^{s,p}_i}, \quad T^{s,p} = \frac{E^{s,p}_t}{E^{s,p}_i} \tag{2}$$

Our main task is to find the coefficients (2) as functions of the incident wave parameters and an arbitrary continuous function $\varepsilon(z)$.

## 2.1. Transmission of a plane wave through an arbitrary one-dimensional layered structure

First, we consider the auxiliary problem of plane wave propagation through a system of $N$ homogeneous layers (shown in Fig. 2) with a dielectric function defined as:

$$\varepsilon(z) = \begin{cases} \varepsilon_0, & z < 0 \\ \sum_{n=1}^{N} \theta(z - z_n + \frac{d_n}{2})\theta(z_n - z + \frac{d_n}{2})\varepsilon_n, & 0 \leq z \leq L \\ \varepsilon_L, & z > 0 \end{cases} \tag{3}$$

Here $\theta(z)$ is the Heaviside step function.

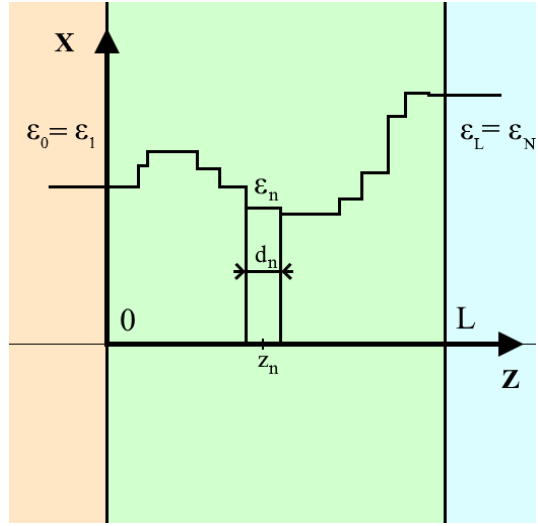

Figure 2. Geometry of the auxiliary problem of plane wave scattering by a layered structure

The scattering amplitudes of the plane wave for structure (3) can be found using the product of $N$ 2×2 matrices [2]

$$\begin{pmatrix} T_N^{s,p} \\ 0 \end{pmatrix} = \prod_{n=1}^{N} \begin{pmatrix} (1/t_n^{s,p})^* & (-r_n^{s,p}/t_n^{s,p})^* \\ -r_n^{s,p}/t_n^{s,p} & 1/t_n^{s,p} \end{pmatrix} \begin{pmatrix} 1 \\ R_N^{s,p} \end{pmatrix} \quad (4)$$

where $T_N^{s,p}$ and $R_N^{s,p}$ are transmission and reflection amplitudes of the whole system of $N$ layers for s- and p-polarizations. Coefficients $t_n^{s,p}$ and $r_n^{s,p}$ are the transmission and reflection amplitudes of the n-th layer, respectively. If each layer is isotropic, has thickness $d_n$ and permittivity $\varepsilon(z) = \varepsilon_n$, we can obtain (based on [21]) the following expressions for the coefficients in Eq. (4):

$$\frac{1}{t_n^s} = \left( \frac{k_{n-1} + k_{n+1}}{2k_{n+1}} \cos[k_n d_n] - i \frac{k_n^2 + k_{n-1} k_{n+1}}{2k_n k_{n+1}} \sin[k_n d_n] \right) C_t \quad (5)$$

$$\frac{r_n^s}{t_n^s} = \left( \frac{k_{n-1} - k_{n+1}}{2k_{n+1}} \cos[k_n d_n] + i \frac{k_n^2 - k_{n-1} k_{n+1}}{2k_n k_{n+1}} \sin[k_n d_n] \right) C_r \quad (6)$$

$$\frac{1}{t_n^p} = \left( \frac{\sqrt{\frac{\varepsilon_{n+1}}{\varepsilon_{n-1}}} k_{n-1} + \sqrt{\frac{\varepsilon_{n-1}}{\varepsilon_{n+1}}} k_{n+1}}{2k_{n+1}} \cos[k_n d_n] - i \frac{\frac{\varepsilon_n}{\sqrt{\varepsilon_{n-1} \varepsilon_{n+1}}} k_n^2 + \frac{\sqrt{\varepsilon_{n-1} \varepsilon_{n+1}}}{\varepsilon_n} k_{n-1} k_{n+1}}{2k_n k_{n+1}} \sin[k_n d_n] \right) C_t \quad (7)$$

$$\frac{r_n^p}{t_n^p} = \left( \frac{\sqrt{\frac{\varepsilon_{n+1}}{\varepsilon_{n-1}}} k_{n-1} - \sqrt{\frac{\varepsilon_{n-1}}{\varepsilon_{n+1}}} k_{n+1}}{2k_{n+1}} \cos[k_n d_n] + i \frac{\frac{\varepsilon_n}{\sqrt{\varepsilon_{n-1} \varepsilon_{n+1}}} k_n^2 - \frac{\sqrt{\varepsilon_{n-1} \varepsilon_{n+1}}}{\varepsilon_n} k_{n-1} k_{n+1}}{2k_n k_{n+1}} \sin[k_n d_n] \right) C_r \quad (8)$$

where $n \in [1, N]$, $C_t = \exp[i(k_{n+1} - k_{n-1}) z_n] \exp[i(k_{n+1} + k_{n-1}) d_n / 2]$, $k_n = (2\pi/\lambda)\sqrt{\varepsilon_n} \cos \alpha_n$, $C_r = \exp[i(k_{n+1} + k_{n-1}) z_n] \exp[i(k_{n+1} - k_{n-1}) d_n / 2]$, $\varepsilon_1 = \varepsilon_0$, $\varepsilon_N = \varepsilon_{N+1} = \varepsilon_L$, $z_n$ defines a point in the middle of the $n$-th layer, $\alpha_n$ is refraction angle of $n$-th layer, which is related to the initial angle of incidence $\alpha$ via the Snell's law $\sqrt{\varepsilon_n}\sin\alpha_n = \sqrt{\varepsilon_0}\sin\alpha$. It is easy to see that, given the condition $k_{n+1} = k_{n-1}$, expressions (5)-(8) will take the well-known form for the case of a homogeneous layer bordering with the same media on both sides.

Calculating the product of $N$ matrices (4) is equivalent to solving a system of finite-difference equations [18]. Thus, the problem can be represented as:

$$D_N^{s,p} = r_N^{s,p}/t_N^{s,p}\,\overline{D}_{N-1}^{s,p} + 1/t_N^{s,p}\,D_{N-1}^{s,p}, \quad N \geq 1 \tag{9}$$

$$\overline{D}_N^{s,p} = \left(1/t_N^{s,p}\right)^{*}\overline{D}_{N-1}^{s,p} + \left(r_N^{s,p}/t_N^{s,p}\right)^{*}D_{N-1}^{s,p}, \quad N \geq 1 \tag{10}$$

Here

$$\begin{aligned} D_N^{s,p} &= 1/T_N^{s,p}, & \overline{D}_N^{s,p} &= (R_N^{s,p}/T_N^{s,p})^{*} \\ D_{N-1}^{s,p} &= 1/T_{N-1}^{s,p}, & \overline{D}_{N-1}^{s,p} &= (R_{N-1}^{s,p}/T_{N-1}^{s,p})^{*} \end{aligned} \tag{11}$$

$T_{N-1}^{s,p}$ and $R_{N-1}^{s,p}$ are the transmission and reflection amplitudes of s- and p-waves for the first $N-1$ layers of the structure (3). To find a particular solution of equations (9) and (10) we set the initial conditions as

$$D_0^{s,p} = 1, \quad \overline{D}_0^{s,p} = 0 \tag{12}$$

which mean that a layer system with zero thickness ($L = 0$) transmits the wave without any change ($T_0 = 1$, $R_0 = 0$).

## 2.2 Transmission of a plane wave through a one-dimensional structure with an arbitrary continuous permittivity

Now we can proceed to the original problem of determining the transmission and reflection amplitudes for a one-dimensional dielectric layer with an arbitrary continuous $\varepsilon(z)$. For this purpose, we introduce the following functions

$$D^{s,p}(z) = 1/T^{s,p}(z), \quad \overline{D}^{s,p}(z) = \left(R^{s,p}(z)/T^{s,p}(z)\right)^{*} \tag{13}$$

here $T^{s,p}(z)$ and $R^{s,p}(z)$ are transmission and reflection amplitudes of the region of the layer located between the points 0 and $z$.

Further, we add a sufficiently thin layer with thickness $2\Delta z$ and constant dielectric permittivity $\varepsilon(z)$ to this region. By applying Eqs. (5)-(8) for this thin layer ($k_n d_n = 2k(z)\Delta z \ll 1$) we obtain

$$\frac{1}{t^s} = 1 - \frac{1}{k}\frac{\Delta k}{\Delta z}\Delta z + 2i\frac{\Delta k}{\Delta z}z\Delta z \tag{14}$$

$$\frac{r^s}{t^s} = -\exp(2ikz)\frac{1}{k}\frac{\Delta k}{\Delta z}\Delta z \tag{15}$$

$$\frac{1}{t^p} = 1 - \frac{1}{k}\frac{\Delta k}{\Delta z}\Delta z + 2i\frac{\Delta k}{\Delta z}z\Delta z \tag{16}$$

$$\frac{r^s}{t^s} = -\cos(2\beta)\exp(2ikz)\frac{1}{k}\frac{\Delta k}{\Delta z}\Delta z \qquad (17)$$

Here $\Delta k = k(z) - k(z-\Delta z) = k(k+\Delta z) - k(z)$, $k(z) = (2\pi/\lambda)\sqrt{\varepsilon(z)}\cos\beta(z)$, $\beta$ is the angle of refraction. One can notice that the expressions for the transmittance coincide for the two polarizations at all angles. This is due to the fact that we have neglected the higher-order terms of $\Delta z$, which are different for the s- and p-waves.

Next, we substitute in Eqs. (9)-(10) the expressions $D_N^{s,p} = D^{s,p}(z+2\Delta z)$, $\overline{D}_N^{s,p} = \overline{D}^{s,p}(z+2\Delta z)$ and $D_{N-1}^{s,p} = D^{s,p}(z)$, $\overline{D}_{N-1}^{s,p} = \overline{D}^{s,p}(z)$, and expand the result into a series of small $\Delta z$. After that, we make $\Delta z \to 0$ and taking into account Eqs. (14)-(17) we obtain the following system of equations for $D^{s,p}(z)$ and $\overline{D}^{s,p}(z)$:

$$\frac{dD^s}{dz} = -\frac{1}{2k}\frac{dk}{dz}\exp(2ikz)\overline{D}^s - \frac{1}{2k}\frac{dk}{dz}D^s + i\frac{dk}{dz}zD^s \qquad (18)$$

$$\frac{d\overline{D}^s}{dz} = -\frac{1}{2k}\frac{dk}{dz}\overline{D}^s - i\frac{dk}{dz}z\overline{D}^s - \frac{1}{2k}\frac{dk}{dz}\exp(-2ikz)D^s \qquad (19)$$

for s-polarization and

$$\frac{dD^p}{dz} = -\frac{\cos(2\beta)}{2k}\frac{dk}{dz}\exp(2ikz)\overline{D}^p - \frac{1}{2k}\frac{dk}{dz}D^p + i\frac{dk}{dz}zD^p \qquad (20)$$

$$\frac{d\overline{D}^p}{dz} = -\frac{1}{2k}\frac{dk}{dz}\overline{D}^p - i\frac{dk}{dz}z\overline{D}^p - \frac{\cos(2\beta)}{2k}\frac{dk}{dz}\exp(-2ikz)D^p \qquad (21)$$

for p-polarization. Here $\frac{dk}{dz} = \frac{2\pi}{\lambda}\frac{1}{\cos\beta}\frac{d\sqrt{\varepsilon}}{dz}$.

The initial conditions for Eqs. (18)-(21) are the following:

$$D^{s,p}(0) = 1, \quad \overline{D}^{s,p}(0) = 0. \qquad (22)$$

Thus, finding the transmission and reflection amplitude of an arbitrarily polarized plane wave reduces to solving a system of linear differential equations (18)-(21) with initial conditions (22). For an arbitrary dependence $\varepsilon(z)$ this system can be solved only numerically.

Further, the system (18)-(22) can be transformed into a more compact and convenient for calculation form. To do so, we introduce the following functions:

$$F^{s,p} = \exp(-ikz)D^{s,p} - \exp(ikz)\overline{D}^{s,p} \qquad (23)$$

$$Q^{s,p} = \exp(-ikz)D^{s,p} + \exp(ikz)\overline{D}^{s,p} \qquad (24)$$

After substituting $F^{s,p}$ and $Q^{s,p}$ into Eqs. (18)-(22), we obtain equations for them:

$$\frac{dF^{s,p}}{dz} = -ikQ^{s,p} - A^{s,p}\frac{1}{k}\frac{dk}{dz}F^{s,p} \qquad (25)$$

$$\frac{dQ^{s,p}}{dz} = -ikF^{s,p} - B^{s,p}\frac{1}{k}\frac{dk}{dz}Q^{s,p} \tag{26}$$

with the initial conditions in $z = 0$:
$$F^{s,p}(0) = 1, \quad Q^{s,p}(0) = 1. \tag{27}$$

Here
$$A^s = 0, \quad B^s = 1, \quad A^p = \sin^2\beta, \quad B^p = \cos^2\beta. \tag{28}$$

By solving the system of differential equations (25)-(26) for complex functions $F^{s,p}$ and $Q^{s,p}$ with initial conditions (27) and finding their values at $z = L$, and also using formulas (13) and (23)-(24), we obtain the reflection and transmission amplitudes of incident light for an inhomogeneous medium of finite thickness $L$:

$$T^{s,p} = \frac{2\exp(-ikL)}{Q^{s,p}(L) + F^{s,p}(L)}, \quad R^{s,p} = \frac{(Q^{s,p}(L) - F^{s,p}(L))^*}{Q^{s,p}(L) + F^{s,p}(L)} \tag{29}$$

In addition, we can easily obtain a formula for the distribution of electric field inside the layer. To do this, let us write down the expression for the electric field at the point $z$ ($0 \leq z \leq L$) similarly to equation (4):

$$\begin{pmatrix} E_+^{s,p}(z) \\ E_-^{s,p}(z) \end{pmatrix} = \begin{pmatrix} (1/T^{s,p}(z))^* & (-R^{s,p}(z)/T^{s,p}(z))^* \\ -R^{s,p}(z)/T^{s,p}(z) & 1/T^{s,p}(z) \end{pmatrix} \begin{pmatrix} 1 \\ R^{s,p}(L) \end{pmatrix} E_i^{s,p} \tag{30}$$

Here $E_i^{s,p}$ is the electric field amplitude of the incident wave at $z = 0$, $E_+^{s,p}(z)$ and $E_-^{s,p}(z)$ are the amplitudes of the waves propagating inside the layer to the right and to the left, respectively. Finally, the total electric field at point $z$ is the sum of $E^{s,p}(z) = E_+^{s,p}(z) + E_-^{s,p}(z)$. By using Eqs. (29)-(30), we obtain the formula for electric field distribution inside the layer:

$$E^{s,p}(z) = \left[(F^{s,p}(z))^* + R^{s,p}(L)F^{s,p}(z)\right]E_i^{s,p} \tag{31}$$

It should be noted that the expressions obtained so far have been written taking into account the absence of radiation losses in the medium. According to the results presented in [23], in order to include absorption (or amplification) correctly in these formulas one should replace the complex conjugation by the wave vector inversion $k \to -k$. Then Eqs. (29) and (31) will take a new form:

$$T^{s,p} = \frac{2\exp(-ikL)}{Q_k^{s,p}(L) + F_k^{s,p}(L)}, \quad R^{s,p} = \frac{Q_{-k}^{s,p}(L) - F_{-k}^{s,p}(L)}{Q_k^{s,p}(L) + F_k^{s,p}(L)} \tag{32}$$

$$E^{s,p}(z) = \left[F_{-k}^{s,p}(z) + R^{s,p}(L)F_k^{s,p}(z)\right]E_i^{s,p} \tag{33}$$

Here, the indices $k$ and $-k$ show the functions calculated before and after the inversion, respectively.

### 3. Application of the new method for various problems

As an illustrative example of applying our proposed method, let us consider the problem of scattering of a plane wave incident at an angle $\alpha$ on a layer with a periodic dependence of dielectric permittivity

$$\varepsilon(z) = \varepsilon_1 + \varepsilon_2 \cos\left(\frac{2\pi}{\Lambda} z\right) \tag{34}$$

where $\varepsilon_1$ is the constant value, $\varepsilon_2$ and $\Lambda$ are the depth and the period of modulation, correspondingly. It is assumed that the layer is sandwiched between two, in general, different media with permittivity $\varepsilon_0 = \varepsilon_1 + \varepsilon_2$ on the left and $\varepsilon_L = \varepsilon_1 + \varepsilon_2 \cos(2\pi L/\Lambda)$ on the right, where $L$ is the layer thickness.

The problem of wave scattering by such a layer has an analytical solution for s-polarization, in which the reflection and transmission coefficients, as well as the amplitude of the electric field inside the medium, are expressed through Mathieu functions [22]. In the case of dependence (34), the Helmholtz equation for the electric field of the s- wave takes the form:

$$\frac{d^2 E_s}{dz^2} + \left(\frac{2\pi}{\lambda}\right)^2 \left(\varepsilon_1 - \varepsilon_0 \sin^2\alpha + \varepsilon_2 \cos\left(\frac{2\pi}{\Lambda} z\right)\right) E_s = 0 \tag{35}$$

It is known that Eq. (35) has a solution $E_s(z) = C_1 ms(a, q, \pi z/\Lambda) + C_2 mc(a, q, \pi z/\Lambda)$, where $ms(a, q, \xi)$ and $mc(a, q, \xi)$ are odd and even Mathieu functions with parameters $a = 4(\varepsilon_1 - \varepsilon_0 \sin^2\alpha)\Lambda^2/\lambda^2$ и $q = -2\varepsilon_2 \Lambda^2/\lambda^2$. The constants $C_1$ and $C_2$, as well as coefficients $R$ and $T$ can be found from the continuity conditions for $E_s$ and $dE_s/dz$ at the boundaries of the layer:

$$1 + R = E_s(0), \quad \frac{dE_s}{dz}(0) = i\frac{2\pi}{\lambda}\sqrt{\varepsilon_0}(R-1)\cos\alpha,$$

$$T = E_s(L), \quad \frac{dE_s}{dz}(L) = i\frac{2\pi}{\lambda}\sqrt{\varepsilon_L} T \cos\gamma, \tag{36}$$

here $\alpha$ is the angle of incidence and $\gamma$ is the angle of refraction. Thus, from the Eq. (36) one can obtain an expression for the reflection and transmission coefficients:

$$R = \frac{p\alpha_{c0}(q\alpha_{sL} + \beta_{sL}) - \beta_{s0}(q\alpha_{cL} + \beta_{cL})}{p\alpha_{c0}(q\alpha_{sL} + \beta_{sL}) + \beta_{s0}(q\alpha_{cL} + \beta_{cL})} \tag{37}$$

$$T = \frac{2p(\alpha_{cL}\beta_{sL} - \alpha_{sL}\beta_{cL})}{p\alpha_{c0}(q\alpha_{sL} + \beta_{sL}) + \beta_{s0}(q\alpha_{cL} + \beta_{cL})} \tag{38}$$

here $\alpha_{sL} = ms(a, q, \pi L/\Lambda)$, $\alpha_{c0} = mc(a, q, 0)$, $\alpha_{cL} = mc(a, q, \pi L/\Lambda)$, $\beta_{s0} = \left.\frac{dms}{d\xi}(a, q, \xi)\right|_{\xi=0}$, $\beta_{sL} = \left.\frac{dms}{d\xi}(a, q, \xi)\right|_{\xi=\pi L/\Lambda}$, $\beta_{cL} = \left.\frac{dmc}{d\xi}(a, q, \xi)\right|_{\xi=\pi L/\Lambda}$, $p = 2i\sqrt{\varepsilon_0}\frac{\Lambda}{\lambda^2}\cos\alpha$, $q = 2i\sqrt{\varepsilon_L}\frac{\Lambda}{\lambda^2}\cos\gamma$. In

deriving Eqs. (37)-(38) we also took advantage of the properties of the Mathieu functions $ms(a,q,0) = 0$ and $\left.\dfrac{dmc}{d\xi}(a,q,\xi)\right|_{\xi=0} = 0$ for any values of $a$ and $q$.

The results of the numerical simulation are shown in Figs. 3-5 together with insets showing the difference between the numerical and the analytical solution. Here and below, the system of equations (25)-(26) was solved by the classical Runge-Kutta method of 4th order with a constant step $h = 5 \times 10^{-4} L$. Figure 3 clearly shows the characteristic wavelength ranges with high reflection coefficient $|R|^2 \approx 1$, which correspond to the photonic band gap (PBG) of a periodic photonic structure. At oblique incidence (Fig. 3(b)), the reflection spectra of the *s*- and *p*- polarizations no longer coincide, and the PBG shifts to the short-wave region as the incidence angle $\alpha$ increases. We can also see that the reflection for the *p*- polarization is smaller throughout the entire spectral range, which is due to $\alpha$ being close to the Brewster angle. The computational error of the reflection coefficient reaches its maxima at the edges of the PBG where the derivative $dR/d\lambda$ is large. The field distribution in Fig. 4 shows an exponential attenuation of the field at a wavelength inside the PBG. The dependence of the reflection coefficient on the thickness of the structure outside the PBG in Fig. 5 exhibits a complex oscillatory behavior associated with multiple wave reflections inside the layer.

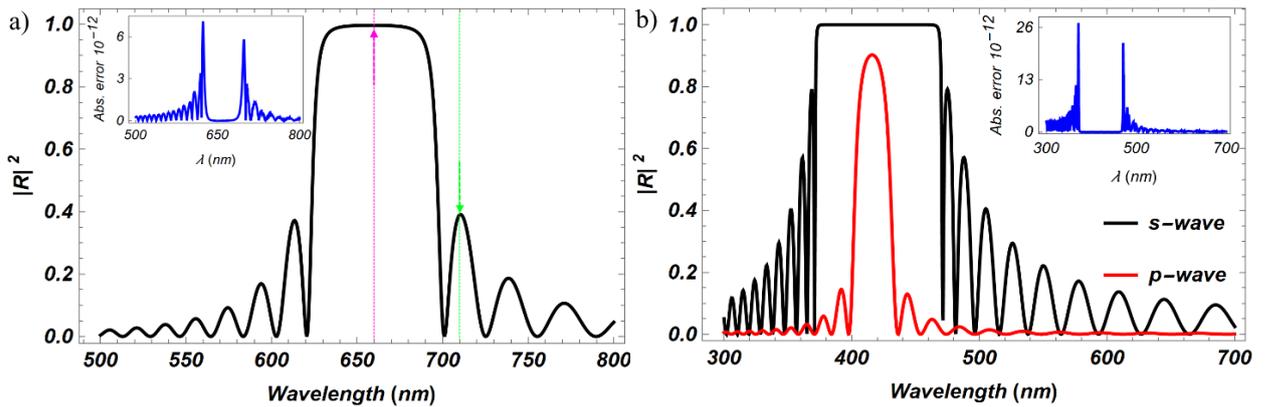

Figure 3. Dependence of the reflection coefficient $|R|^2$ on the wavelength (a) at normal incidence, (b) at incidence angle $\alpha = 45°$, as well as the insets with the absolute difference from the exact formula (37). Layer parameters: $\Lambda = 200$ nm, $L/\Lambda = 25$, $\varepsilon_1 = 2.723$, $\varepsilon_2 = 0.5$.

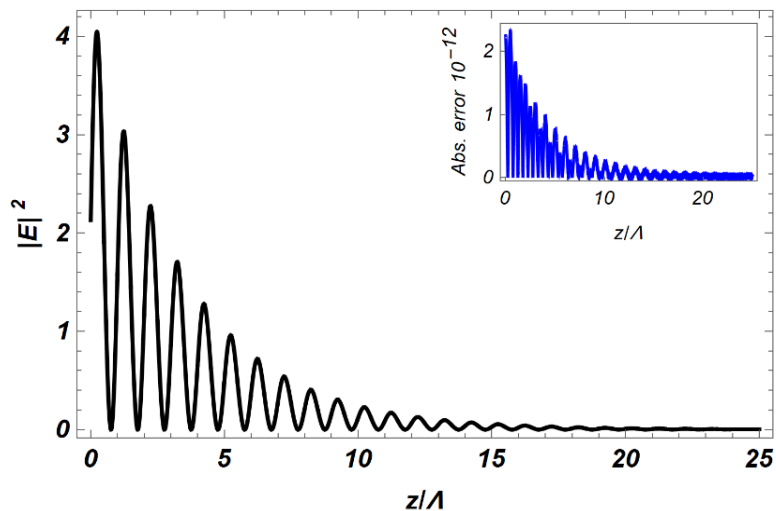

Figure 4. Intensity of the electric field inside the layer at $\lambda = 660$ nm (shown by the pink arrow in Fig. 3) at normal incidence, as well as the inset with the absolute difference from the exact formula. All parameters are the same as in Fig. 3.

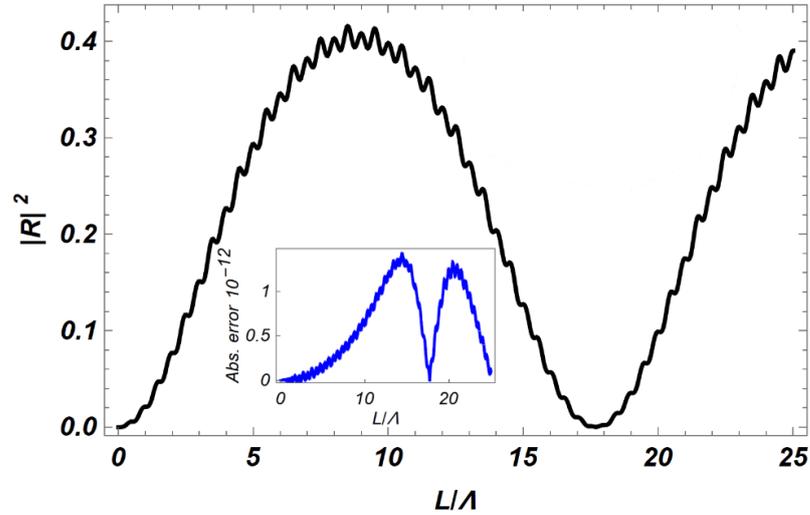

Figure 5. Dependence of the reflection coefficient $|R|^2$ on the layer thickness at $\lambda = 710$ nm (shown by the green arrow in Fig. 3) at normal incidence, as well as the inset with the absolute difference from the exact formula (37). All parameters are the same as in Fig. 3.

Now, as an example of using the new method for aperiodic structures, we will proceed to wave scattering by a layer with gradient parameters. In the first case, the layer has dielectric permittivity (34), but now with a modulation period gradient (chirped grating):

$$\Lambda(z) = \Lambda_0 + \delta\Lambda \exp(-\kappa z / L) \tag{39}$$

Figure 6(a) shows the calculated reflection spectrum, which shows a considerable asymmetry of the spectrum with respect to the PBG, which is expressed in a greater reflection on the long-wave side. This is due to the reflection of the wave from the parts of the layer with a longer modulation period. Figure 6(b) shows a comparison of the field inside the layer when the radiation is incident from the left and right side of the layer. An interesting fact is that, despite the coincidence of the reflection and transmission spectra of such a layer for both directions, the field distribution inside the layer can have a significant asymmetry. Figure 7 shows the angular dependence of the reflection spectrum for the two polarizations. The angular range is restricted by the angle of total internal reflection at which $\cos\beta$ in the denominator of Eqs. (25)-(26) turns to zero.

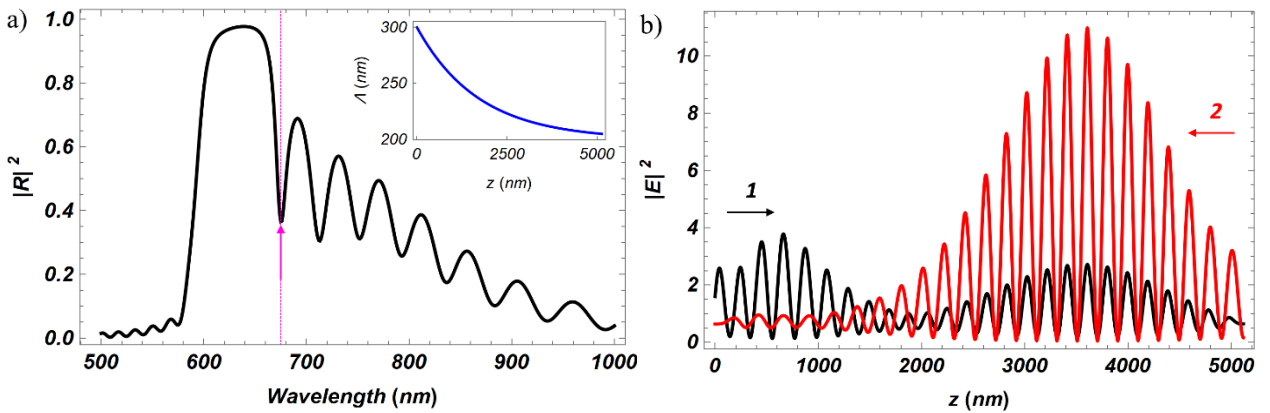

Figure 6. (a) Wavelength dependence of the reflection coefficient $|R|^2$ at normal incidence, as well as the inset with the permittivity modulation period dependence; (b) field intensity distribution inside the layer at $\lambda = 675$ nm (shown by the pink arrow) in 1) the forward direction and 2) the backward direction. The layer parameters are $\Lambda_0 = 200$ nm, $\delta\Lambda = 200$ nm, $\kappa = 3$, $\varepsilon_1 = 2.723$, $\varepsilon_2 = 0.5$. The layer is $L = 5117$ nm thick and contains 25 oscillations of dielectric permittivity.

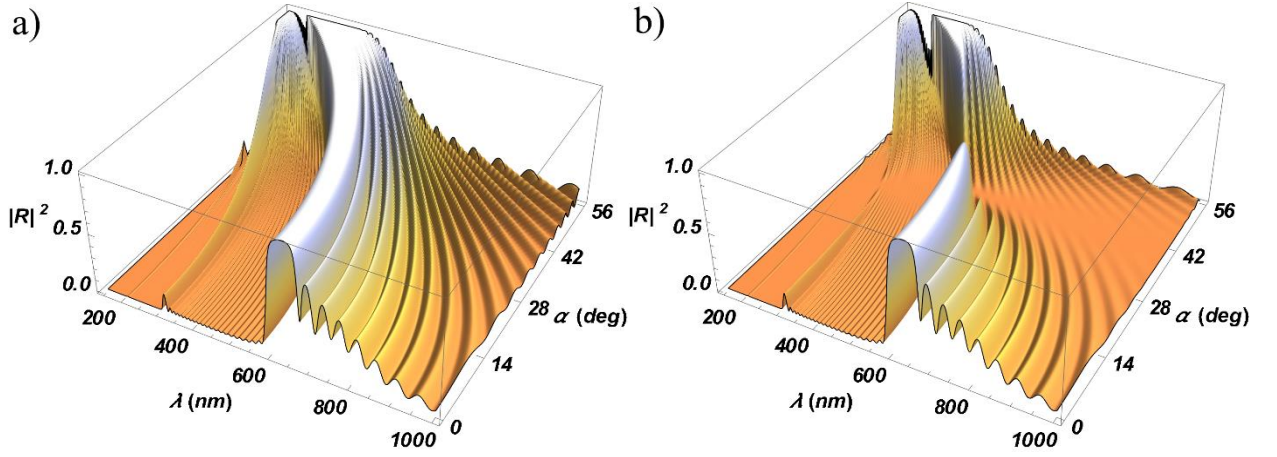

Figure 7. Angular dependence of the reflection spectrum of the chirped layer (a) for *s*- polarization and (b) for *p*- polarization. All parameters are the same as in Fig. 6.

In the second case, the layer also has dielectric permittivity (34), but now with a modulation amplitude gradient (apodized grating):

$$\varepsilon_2(z) = \varepsilon_{20} + \delta\varepsilon_2 \exp(-\kappa z / L) \qquad (40)$$

Figure 8(a) shows the calculated reflection spectrum. In this case, the spectrum is mostly symmetric with respect to the PBG and differs from the case of a strictly periodic lattice by the larger reflection at the edges of the PBG. Also, similar to the case of the chirped grating, we have an asymmetric field distribution when the direction of wave propagation changes (Fig. 8(b)). Figure 9 shows the angular dependence of the reflection spectrum for the two polarizations.

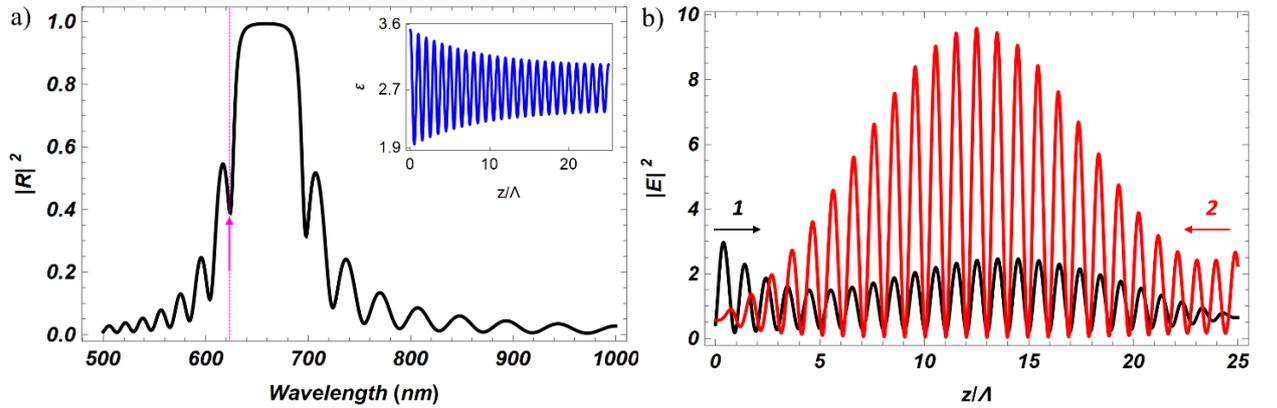

Figure 8. (a) Dependence of the reflection coefficient $|R|^2$ on the wavelength at normal incidence, as well as the inset with the dependence of the dielectric permittivity along the layer; (b) Distribution of the field intensity inside the layer at $\lambda = 623$ nm (shown by the pink arrow) in 1) the forward direction and 2) the reverse direction. The parameters of the layer $\varepsilon_{20} = 0.3$, $\delta\varepsilon_2 = 0.5$, $\kappa = 3$, $\varepsilon_1 = 2.723$, $\Lambda = 200$ nm, $L/\Lambda = 25$.

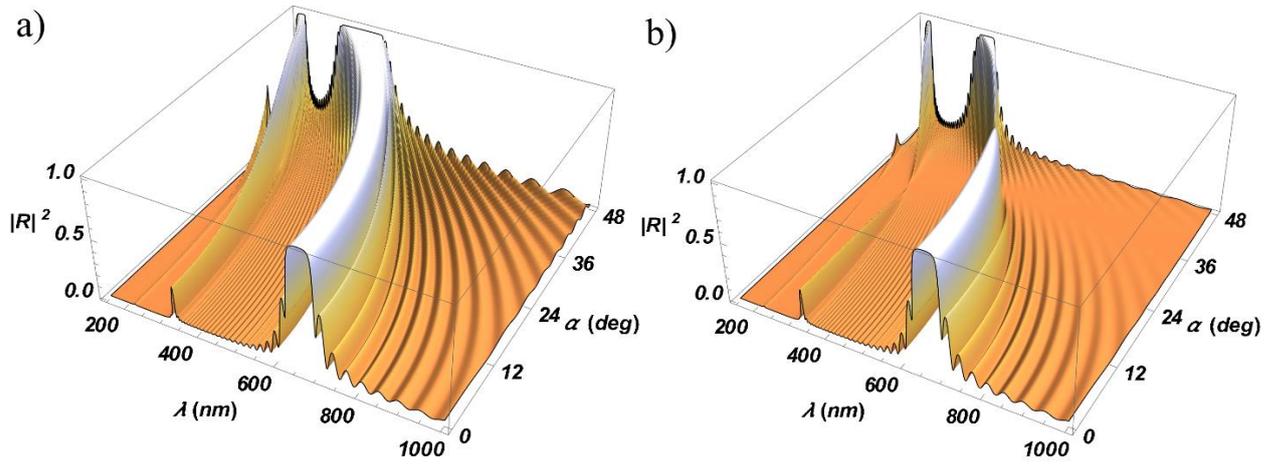

Figure 9. Angular dependence of the reflection spectrum of the apodized layer (a) for *s*- polarization and (b) for *p*- polarization. All parameters are the same as in Fig. 8.

Finally, let us consider transmission of a wave through a plasma layer with harmonic modulation of the electron concentration:

$$N_e = N_0 \left(1 - \cos(2\pi z / \Lambda)\right) \quad (41)$$

It is well known that the plasma permittivity depends on the frequency of the incident radiation and has the following form:

$$\varepsilon = 1 - \frac{\omega_p^2}{\omega(\omega + i\eta)} \quad (42)$$

where $\omega_p = \sqrt{4\pi e^2 N_e / m_e}$ is plasma frequency, $e$ and $m_e$ are the charge and the mass of electron, $\eta$ is electron collision frequency. Since in this case there is an imaginary part of dielectric permittivity, which is responsible for absorption, it is now necessary to use formulas (32)-(33). Figure 10 shows the obtained reflection and absorption spectra of the plasma layer, from which one can see two PBGs at wavelengths of 6.5 mm and 3.7 mm. Within the PBG, a minimum of absorption is observed, due to the strong reflection from the periodic structure and short effective penetration depth. The strong increase in reflection at $\lambda > 10$ mm is not due to resonance scattering by the periodic structure, but due to the presence of layer regions with negative $Re(\varepsilon)$. In our case, such regions appear at wavelengths greater than 7.5 mm, and continue to grow as the incident wavelength increases, leading to an increase of the reflection from the layer. At wavelengths less than 3.5 mm $\varepsilon$ is barely different from unity, so the transmittance begins to dominate over reflection and absorption. Figure 11 shows the field distribution at a wavelength of 7.2 mm where $Re(\varepsilon) > 0$ and attenuation is due to absorption and reflection from the periodic structure, and at a wavelength of 10 mm, when attenuation is due to absorption and reflection from areas with $Re(\varepsilon) < 0$.

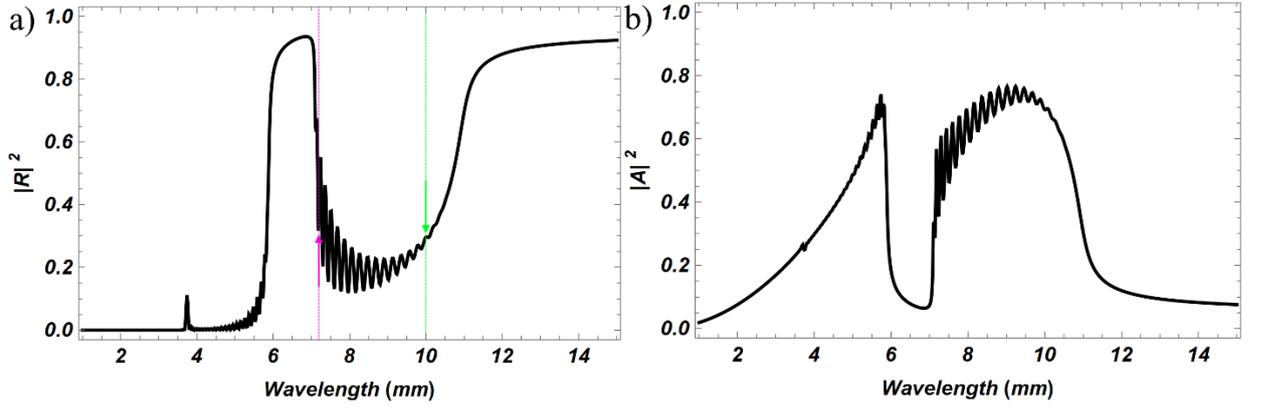

Figure 10. (a) Reflection $|R|^2$ and (b) absorption $|A|^2 = 1 - |R|^2 - |T|^2$ spectra of the plasma layer at normal incidence. Layer parameters: $\Lambda = 4$ mm, $L/\Lambda = 25$, $N_0 = 10^{13}$ cm$^{-3}$, $\gamma = 0.03\omega_p$.

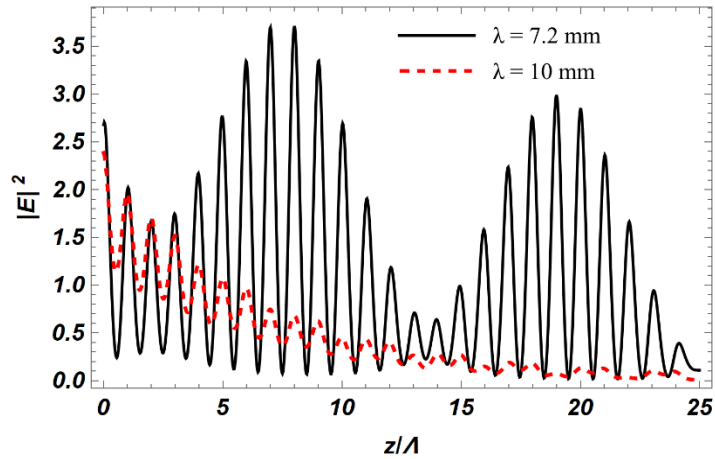

Figure 11: Field intensity distribution inside the plasma layer at two wavelengths (shown by arrows in Fig. 10) at normal incidence. All parameters are the same as in Fig. 10.

### 4. Conclusions

To sum up, we have developed a new method for determining the reflection and transmittance coefficients for oblique incidence of a plane electromagnetic wave on an inhomogeneous isotropic layer of finite thickness. The method is based on the substitution of the boundary problem for the wave equation into a Cauchy problem for a system of two first-order differential equations. This is achieved by a suitable choice of unknown functions $F$ and $Q$ representing a combination of reflection and transmission coefficients of the layer. From a computational point of view, this method compares favorably with standard methods in that the Cauchy problem for the system of equations is easier to solve than the boundary problem for the wave equation. This method considers the reflection and transmittance through each elementary layer with respect to neighboring layers rather than with respect to vacuum as in the method [18], which leads to two main differences: first, the system of equations on the unknown functions and contains a derivative of the refractive index (or permittivity), which can be used, for example, when looking for analytical solution; second, instead of fixed and equal on both sides of the layer, the refractive indices of the external media are bound to the values at the boundaries of the layer by the continuity condition, and therefore are not necessarily equal to each other, which simplifies the calculation for the case of different media on both sides of the layer. Also, this method makes it easy to calculate the field distribution inside the layer. The main drawback of the new method, however, is the requirement of continuity of the refractive index over the entire layer thickness, although we plan to continue work in the future to get around this restriction. The reflection spectra and field distribution obtained using this method were compared with the analytical solution based on Mathieu functions.

## Acknowledgments

The work was supported by the Foundation for the Advancement of Theoretical Physics and Mathematics "BASIS" (Grant № 21-1-1-6-1).

## Conflict of interest

No potential conflict of interest was reported by the authors.